\documentclass[%
 reprint,
nofootinbib,
bibnotes,
 amsmath,amssymb,
]{revtex4-2}
\usepackage[usenames]{color}
\usepackage {graphicx}
\usepackage{bm}

\begin{document}

\title{Consequences of a Two-Time Relativistic Bohmian Model}

\author{Giuseppe Ragun{\'i}}
\email{giuseppe.raguni@um.es , giuseraguni@gmail.com}
\affiliation{{Departamento de Física, Universidad de Murcia, E-30071 Murcia, Spain.\\
\rm {Preprint of an article published in \emph{Communications in Theoretical Physics} on June 20, 2024 DOI: 10.1088/1572-9494/ad48fe}}}


\begin{abstract}
Effects of a Bohmian type quantum-relativistic theory are explored. The model is obtained by introducing a new and independent time parameter whose relative motions are not directly observable and cause the quantum uncertainties of the physical observables. Unlike the usual de Broglie-Bohm theories, the Quantum Potential does not affect directly the observable motion but determines the one that is relative to the new time variable. It turns out that the \emph{Zitterbewegung} of a free particle, of which a more general law is obtained, is the key example of these hidden motions and, through it, it seems possible to give physical reality to the Feynman's paths. A relativistic revision of the uncertainty principle also derives from the theory.

\medskip{}
\end{abstract}


\maketitle

\section{Introduction}
At today, while the de Broglie-Bohm (or pilot-waves) theory  \cite{01,02,03,04,05,06} enjoys some favor as a deterministic approach to non-relativistic Quantum Mechanics, the picture is not so favorable as regards the attempts at its relativistic generalization. For some time, the characteristics of the instantaneous action at a distance admitted by the theory has been considered a manifest violation of the Lorentz invariance \cite{07,08,09,10}. Indeed, it has been shown that this problem can be solved, but at the cost of admitting superluminality \cite{11,12,13,14,15}.
Currently, the most authoritative Lorentz-covariant pilot waves theories can be grouped in this way, in terms of the proposed strategy to define simultaneity \cite{16}:

\begin{itemize}
\renewcommand{\labelitemi}{$-$}
\item Introducing a temporal parameterization that synchronizes the trajectories
of all particles \cite{11,12,13,17,18}.
This approach has the drawback of not making concrete statistical
predictions, although subsequent developments suggest this possibility
under certain circumstances \cite{15,19,20,21}.
\item Defining a future light-cone as surface of simultaneity \cite{22}.
This theory is unable to predict any probability.
\item Rejecting the decomposition of the wavefunction into norm and phase, and interpreting it as a real quantum field exerting force on the
particle \cite{23}. Here, tachyon particles are avoided but it is not clear if and how the model can be verified.
\item Defining simultaneity through "slicings"
of the space-time (\emph{foliations}) by spacelike hypersurfaces.
The majority of researchers favor a preferred foliation, unique for
all particles \cite{24,25,26,27,28,30} but there are also those who suggest a degenerate foliation \cite{29} or an union of all possible foliations \cite{31}. 
\end{itemize}

The theories of the last group are thought by some researchers to be the most promising way to obtain the standard statistical predictions \cite{32,33}. However, these models introduce additional spacetime structures and admit 
superluminal velocities (and hence travels backwards
in time in some Lorentz frames) for massive particles, in violation of Special Relativity, although not of Lorentz covariance. A more general reason for dissatisfaction is the fact that all the mentioned theories only aim to deterministically reproduce the predictions of standard quantum mechanics; for an advantage, therefore, exclusively conceptual, superfluous from a strictly phenomenological point of view. On the other hand, there is no shortage of fundamental quantum topics, for example spin and \emph{Zitterbewegung} \cite{34,35}, waiting to be satisfactorily understood and explained.
It would certainly be desirable to have a relativistic quantum theory that is able to propose, together with a deterministic ontology, predictions \emph{beyond} the standard interpretation, possibly in full compliance with the spirit of Special Relativity.

In this article, after observing that the more spontaneous relativistic generalization of the original de Broglie-Bohm theory is unsatisfactory as regards the continuity of the Quantum Potential with the non-relativistic case, we explore the strategy of introducing a new independent time parameter as unique hidden variable, whose relative motions are recognized to be instantaneous with respect to the traditional time. In this deterministic and explicitly non-local approach, a more general \emph{Zitterbewegung}, as motion in the new time variable caused by the Quantum Potential, arises; then, it would be not directly observable, in accordance with recent beliefs \cite {36,37,38}. The model is potentially  capable to assign full physical reality to the Feynman paths and predicts, as non-standard fact, a modification of the uncertainty principle: by depending on the direction of motion, it would become anisotropic. This property could be observed in high energy accelerators.

\section{Relativistic generalization issue for the \emph{quantum potential} in case of constant momentum}
In the de Broglie-Bohm non-relativistic theory  \cite{01,02,03,04,05,06}, the wavefunction:
\begin{equation}
\psi(\vec{r},t)=R(\vec{r},t)\,{\rm{e}}^{\frac{\rm {i}}{\hbar}S(\vec{r},t)},
\end{equation}
with \emph{$R>0$} and \emph{S }real functions, is inserted into the
Schr{\"o}dinger equation, yielding the following equations:
\begin{equation}
\frac{\partial S}{\partial t}+\frac{(\vec{\nabla}S){}^{2}}{2m}\,+V-\frac{\hbar{}^{2}}{2m}\,\frac{\nabla{}^{2}R}{R}=0,
\end{equation}
\begin{equation}
\frac{\partial R^{2}}{\partial t}+\vec{\nabla}\cdot(R^{2}\frac{\vec{\nabla}S}{m})=0.
\end{equation}
Identifying \emph{S} with the Hamilton's principal function, we
have $\vec{\nabla}S=\vec{p}$ and $\frac{\partial S}{\partial t}=-H$, so eq. (2) means the Hamilton-Jacobi equation with Hamiltonian:
\begin{equation}
H=\frac{p{}^{2}}{2m}+V-\frac{\hbar{}^{2}}{2m}\,\frac{\nabla{}^{2}R}{R},
\end{equation}
where the last addend represents an energy, called \emph{Quantum
Potential}, that takes into account the quantum effects:
\begin{equation}
V_{Q}\equiv-\frac{\hbar{}^{2}}{2m}\,\frac{\nabla{}^{2}R}{R}.
\end{equation}
According to the Bohm's original proposal, the motion equation is:
\begin{equation}
\frac{{\rm{d}}  \vec{p}}{{\rm{d}} t}=-\vec{\nabla}V-\vec{\nabla}V_{Q}.
\end{equation}

In the case of a free particle (\emph{V}=0), both a solution with variable $V_{Q}$ and a stationary solution with positive constant $V_{Q}$ are admitted \cite{39,40}.

The direct relativistic generalization of this approach, still for a free particle, pushes to
write: $\vec{\nabla}S=\vec{p}=m\,\gamma\,\vec{v}$ and $\frac{\partial S}{\partial t}=-H=-m\,\gamma\,c^{2}-V_{Q}$, with $V_{Q}$ to be determined. We will start by assuming that
$\vec{p}$ is constant, while for $V_{Q}$, and therefore \emph{H},
a possible dependence on time.

From eq. (1), taking the gradient, then the divergence and dividing by $\psi$, we obtain:
\begin{equation}
\frac{\nabla{}^{2}\psi}{\psi}=\frac{\nabla{}^{2}R}{R}+2\,\frac{\rm{i}}{\hbar}\,\frac{\vec{\nabla}R}{R}\,\cdot\,\vec{p}-\frac{p^{2}}{\hbar{}^{2}},
\end{equation}
being $\vec{\nabla}\cdot\vec{v}=0$. Comparison with the Klein-Gordon
equation:
\begin{equation}
\nabla{}^{2}\psi-\frac{1}{c^{2}}\,\frac{\partial^{2}\psi}{\partial t^{2}}=\frac{m{}^{2}c^{2}}{\hbar{}^{2}}\,\psi,
\end{equation}
provides:
\begin{equation}
\frac{\nabla{}^{2}R}{R}+2\,\frac{\rm{i}}{\hbar}\,\frac{\vec{\nabla}R}{R}\,\cdot\,\vec{p}=\frac{m{}^{2}\gamma^{2}c^{2}}{\hbar{}^{2}}+\frac{1}{c^{2}\psi}\,\frac{\partial^{2}\psi}{\partial t^{2}},
\end{equation}
having applied the identity $p^{2}+m^{2}c^{2}=m^{2}\gamma^{2}c^{2}$.

On the other hand, deriving twice eq. (1) over time, one gets:
\begin{equation}
\frac{1}{\psi}\frac{\partial^{2}\psi}{\partial t^{2}}=\frac{1}{R}\frac{\partial^{2}R}{\partial t^{2}}-2\,\frac{\rm{i}}{\hbar}\,\frac{H}{R}\frac{\partial R}{\partial t}-\frac{H^{2}}{\hbar{}^{2}}-\frac{\rm{i}}{\hbar}\frac{\partial H}{\partial t}.
\end{equation}
Substituting eq. (10) into (9) and matching real and imaginary parts, we have:
\begin{equation}
H^{2}=m^{2}\gamma^{2}c^{4}+\hbar{}^{2}(\frac{1}{R}\frac{\partial^{2}R}{\partial t^{2}}-c^{2}\frac{\nabla{}^{2}R}{R}),
\end{equation}
\begin{equation}
c^{2}\frac{\vec{\nabla}R}{R}\cdot\vec{p}+\frac{H}{R}\,\frac{\partial R}{\partial t}+\frac{1}{2}\frac{\partial H}{\partial t}=0.
\end{equation}

Now we assume that eq. (3), with $m\gamma$
instead of \emph{m}, still holds (we will justify later this continuity equation). By employing $\vec{\nabla}\cdot\vec{v}=0$, we get just the expansion of $\frac{{\rm{d}}R^{2}}{{\rm{d}}t}$. Of course, we can refer it simply to  \emph{R}:
\begin{equation}
\frac{\partial R}{\partial t}+\vec{v}\cdot\vec{\nabla}R=\frac{{\rm{d}}R}{{\rm{d}}t}=0,
\end{equation}
as well as to any function of \emph{R}. Indeed, since $\vec{v}$ is constant, eq. (13) also holds for any function of any-order derivative of $R$ over time or space, and thus for $V_{Q}$ and \emph{H}. We are therefore assuming $\frac{{\rm{d}}H}{{\rm{d}}t}=0$, but with the possibility that $\frac{\partial H}{\partial t}\neq0$, and the same for $V_{Q}$. Substituting eq. (13) into eq. (12) one finds:
\begin{equation}
\frac{1}{R}\frac{\partial R}{\partial t}(H-m\gamma c^{2})+\frac{1}{2}\frac{\partial H}{\partial t}=0,
\end{equation}
where, looking for non-trivial solutions, one assumes \emph{$m\gamma\neq0$}, and $\frac{\partial R}{\partial t}\neq0$. 

The issue, now, is that equation of motion (6) implies a too drastic simplification of eq. (14). Indeed, it reads:
 
\begin{equation}
\frac{{\rm{d}}\vec{p}}{{\rm{d}}t}=-\vec{\nabla}V_{Q}=0,
\end{equation}
then, applying eq. (13) to $V_{Q}$, one obtains $\frac{\partial V_{Q}}{\partial t}=0$ and so it is also
$\frac{\partial H}{\partial t}=0$. Therefore eq. (14) simply reduces to $H=m\,\gamma\,c^{2}$,  finally resulting $V_{Q}=0$.

So, in the assumptions made, we have not general continuity for
the Quantum Potential with the non-relativistic case.

\section{The two-time Bohmian model}
In the first place it is therefore necessary to admit that, even in the case of a free particle, $\vec{p}$ is not constant. Indeed, inside the Heisenberg \emph{picture} of the standard interpretation, Schr{\"o}dinger and Dirac recognized that an \emph{omnidirectional } velocity - which we will call  $v_i$, with $i$ standing for \emph{intrinsic} - of a sinusoidal motion of frequency $\frac{2m\gamma c{}^{2}}{\hbar}\,$ and amplitude about
$\frac{\hbar}{2m\gamma c}\,$, known as \emph{Zitterbewegung}, must be superimposed on the  actually observed velocity in average value \cite{34,35}. This motion has been much commented on, sometimes related to spin \cite{41,42,43,44} and Quantum Potential \cite{45}, but the debate on its exact interpretation, and even observability, does not seem yet resolved \cite{36,37,38,46,47,48,49,50,51,52,53,54,55}. Hopefully, to rediscover this peculiar movement inside a deterministic theory could clarify its nature. It can only be due to the Quantum Potential; but equation of motion (6) still seems inadequate: in fact it has been shown that putting
$\frac{{\rm{d}}p_i}{{\rm{d}}t}=-\mid\vec{\nabla}V_{Q}\mid$ leads to inconsistency in {[}1 + 1{]} dimensions \cite{40}.

Here we explore the effect of introducing a new independent time variable for the spatial coordinates of the particle. {This proposal certainly demands some justification. The Feynman's sum over histories interpretation of Quantum Mechanics (compatible with the standard one) actually operates \emph{as if}  matter could reside simultaneously in multiple places in space. If we assume this "\emph{as if}" really be a "\emph{so it is}", then the problem arises of explaining how this can happen. An object can occupy different positions in space thanks to movement, which is parameterized by the usual time. If we imagine that there exists an extra time independent of the latter - that is, free to flow while the usual time is \emph{frozen} -, then there may be a hidden movement capable of assigning to the object different positions, which so will appear \emph{simultaneous} with respect to the usual time. It is of fundamental importance to keep in mind that \emph{this simultaneity is totally different from the one that  can exist between two $t$-events}, which is considered (and rightly relativised) in Special Relativity. Here the recognized hidden motion occurs with respect to a new and independent temporal parameter and therefore, as long as the object does not interact with something to constitute a $t$-event, it \emph{must be considered as simultaneous in any reference system}. Section V will be dedicated to the in-depth analysis of this phenomenology.}

Two time physics has been considered in various contexts. An additional temporal parameter is considered e.g. in \cite{56,57,58,59,60} and, with a different approach, in \cite{61,62,63,64} in order to more accurately explain different quantum properties or obtain supersymmetries. But in these theories the new time is not independent from the old one as we wish.

Let's admit, then, that the coordinates ($X, Y, Z$) of the particle depend not only on commonly experienced time \emph{t}, but also on another, totally independent and hidden, time parameter
$\tau$. 
Referring for simplicity only to \emph{$X(t,\tau)$}, we get two velocities, one \emph{classic} and one \emph{intrinsic}: 
\begin{equation}
v_{c_x}\equiv\frac{\partial X}{\partial t}\quad\quad v_{i_x}\equiv\frac{\partial X}{\partial\tau},
\end{equation}
having so: ${\rm{d}}X=v_{c_x}{\rm{d}}t+v_{i_x}{\rm{d}}\tau$. If we suppose that $v_{c_x}$ and $v_{i_x}$ depend, respectively, only
on $t$ and $\tau$, we have:
\begin{equation}
X(t,\tau)=X_{t_0}+X_{\tau_0}+\intop_{0}^{t}v_{c_x}(\xi)\,{\rm{d}}\xi+\intop_{0}^{\tau}v_{i_x}(\zeta)\,{\rm{d}}\zeta,
\end{equation}
having set $t_0=\tau_0=0$.

The intrinsic motion is not directly observable and, for the independence of 
$\tau$, is perceived as instantaneous with respect to time \emph{t}. This explicitly non-local behavior has  dramatic effects on the measurements of the observable quantities. As for the position, eq. (17) allows to the particle, observed in its trajectory as a function of \emph{t},
to jump instantly from one point of space to another, in principle
arbitrarily distant, through the independent time dimension $\tau$; and this has to be generalized in every direction of space. Actually, Quantum Mechanics has accustomed us to even more surprising facts, not limited to the cases of an electron spreaded in an atomic orbital, quantum diffraction or entanglement, but really in \emph{any motion}. Indeed, according to Feynman's \emph{sum over histories} interpretation (which is equivalent to the \emph{standard} one) the physical event "the particle (or photon) moves from \emph{A} to \emph{B}" is only correctly explained by admitting that it has traveled \emph{at once} all the possible paths from \emph{A} to \emph{B}, also interacting with every possible virtual particle of the vacuum \cite{65}. And this is excellently confirmed by experience.

{Let's continue now to specify the dynamics of the model keeping in mind that, due to its omnidirectionality, the intrinsic velocity cannot be treated as a vector}.

For the generic particle wavefunction:
\begin{equation}
\psi(\vec{r},t,\tau)=R(\vec{r},t,\tau)\,{\rm{e}}^{\frac{\rm{i}}{\hbar}S(\vec{r},t,\tau)},
\end{equation}
we admit, justifying it later, a two-time action $S=S_t+S_\tau$ so defined:

$S_t=\int{(-\frac{mc^2}{\gamma_c}-V-V_Q){\rm{d}}t},\
S_{\tau}=\int(\frac{1}{2}m\gamma_{c} v_{i_s}^{2}-V_{Q}){\rm{d}}\tau$, {where $v_{i_s}$ is the intrinsic velocity in the generic direction $\hat s$. Here we have to point out in advance that, on respect to two different spatial directions $\hat{s}$ and $\hat{u}$,
it will result
$v_{i_s}\neq v_{i_u}$. Therefore $\psi$  depends on $s$ through $S_\tau$ (it is actually a $\psi_s$) and hence   the particle comes to be represented with a different wavefunction for each direction of space. Of course, we will demand that the meaning of $\psi$, and in particular of $R$, be the same as in the standard case.}

The motion is decomposed into a classical and a purely quantum part.
The law of classical motion is well known; in particular, if \emph{V} is conservative, the \emph{classic} total energy
is constant:
\begin{equation}
m\gamma_{c}c^{2}+V=const\equiv E_{c},
\end{equation}
and deriving over time one can get:
\begin{equation}
\frac{{\rm{d}}\vec{p}_{c}}{{\rm{d}}t}=-\vec{\nabla}V,
\end{equation}
where the Quantum Potential does not appear; that is also true, naturally, for its non-relativistic approximation. \emph{This equation is therefore  incompatible with the Bohm's motion equation} (6). On the other hand, if we take the total derivative over $t$ of eq. (4), we re-obtain the eq. (20), because $H-V_Q=E_c=const$. \emph{The model is thus in disagreement also with the traditional non-relativistic de Broglie-Bohm theory}. Nevertheless, due to its setting,  we will continue to call it  \emph{Bohmian}.

For $t$-motion we also require the continuity equation:
\begin{equation}
\frac{\partial R^{2}}{\partial t}+\vec{\nabla}\cdot(R^{2}{\vec{v_c})}=0.
\end{equation}
{It expresses the conservation of the probability of \emph{finding the particle} in a unit of volume and its validity, undisputed in the non-relativistic case, must extend to the relativistic one. In fact, this probability is independent of the relativistic mass of the particle and could depend only on its Lorentzian dimensional contraction, which is neglected when treating the particle as a material point.}

As regards the intrinsic motions, it is spontaneous to consider that they are \emph{the direct cause of the quantum uncertainties.} {
However, if this is assumed, then \emph{we must exclude that Special Relativity holds for them.} We will conclude this in the next section, studying the case of the free particle. For now, we just observe that
this condition is in agreement  with the standard \emph{Zitterbewegung},
where the mass of the particle does not undergo any relativistic increase
- other than that one due to the classic velocity $\vec{v_c}$ - despite it
reaches the speed \emph{c}}.

If this is assumed, we  have: $p_{i_s}=m\gamma_{c}v_{i_s}$, which justifies the previous definition of $S_\tau$. 

For the generic $\tau$-motion we also have to suppose the validity of:
\begin{equation}
\frac{\partial R}{\partial \tau}-v_{i_s}\frac{\partial R}{\partial s}=0,
\end{equation}
for any $\hat s$. {It establishes that $R$ is a regressive wave, which moves in a specular manner with respect to the particle. We will provide a justification of this key equation in the Appendix.}

Since $\tau$-motions are not directly influenced by $V$, we can consider that $v_{i_s}$ only depends on $\tau$; consequently, eq. (22) holds for any function of any-order derivative of $R$ over $t$ or space. Hence we are admitting that it can be applied to $V_Q$, at least in case of conservative $V$.

Conservation of energy for the generic $\tau$-motion  reads:
\begin{equation}
V_{Q}+\frac{1}{2}m\gamma_{c}v_{i_s}^{2}\equiv E_{q_s},
\end{equation}
representing a \emph{purely quantum} total energy, independent explicitally on
$\tau$.
Taking the $\frac{\partial}{\partial\tau}$ of eq. (23) and using eq. (22) for  $ V_{Q}$, one gets:
\begin{equation}
m\gamma_{c}\frac{{\rm{d}}v_{i_s}}{{\rm{d}}\tau}=-\frac{\partial V_Q}{\partial s},
\end{equation}
for every $\hat s$. Eqs. (20) and (24) have so to replace the Bohm' motion eq. (6).
\medskip{}

The new Bohmian model can thus be so summarized: 
\begin{enumerate}
\item The spatial coordinates of the particle have to vary as a function
of two independent temporal parameters, \emph{t }and $\tau$. Motions in $\tau$ occur in all directions; they are responsible for
quantum uncertainties and not directly observable.
\item The particle is represented by the wavefunction (18), where $\frac{\partial S}{\partial t}=\frac{\partial S_t}{\partial t}=-H=-m\gamma_{c}c^{2}-V-V_Q$ and $\vec{\nabla}S=\vec{\nabla}S_t=\vec{p}_{c}=m\gamma_{c}\vec{v_c}$. It obeys the generalized (standard replacements: ${\rm{i}}\hbar\frac{\partial}{\partial t}$$\:\rightarrow$$\:{\rm{i}}\hbar\frac{\partial}{\partial t}-V$ and $-{\rm{i}}\hbar\vec{\nabla}$$\:\rightarrow$$\:-{\rm{i}}\hbar\vec{\nabla}-\vec{P}$) Klein-Gordon equation. 
\item For $t$-motion and each $\tau$-motion the continuity eq. (21) and the wave eq. (22), respectively, hold.
\item The Bohm's law
of motion (6) is replaced by eqs. (20) and (24) to get, respectively, \emph{t}-motion and generic $\tau$-motion; for the latter the Special Relativity
is not valid.
 \end{enumerate}
Certainly, guidance equation $\vec{p}_{c}=\hbar\,{\rm{Im}}\,(\frac{\vec{\nabla}\psi}{\psi})$ is still obtained by taking the gradient of $\psi$ and applying $\vec{\nabla}S=\vec{p}_{c}$.

It should also be noted that, although each $\tau$-motion does not respect the Special Relativity, the maximum value of $v_{i_s}$ is $c$: in fact it represents the quantum uncertainty on the generic  component of $\vec {v_c}$, which cannot be greater than $c$.
\medskip{}

The model therefore adds a Newtonian time to the usual four-dimensional $t$-spacetime. As we will see in the next sections, the motions in the new time variable are absolutely negligible for macroscopic objects.
Conversely, when this is not the case, it happens that, even assuming to know the Quantum Potential with a good approximation, due to the absolute
independence of the two times and the omnidirectionality of the $\tau$-motions, the function $\vec{r}(t,\tau)$ is unable to provide us the actual trajectory followed by the particle: it rather represents a three-dimensional object of $t$-spacetime. In particular, while $t$ remains constant, we will find that the particle can $\tau$-oscillate an arbitrary number of times in any arbitrary direction. However, this is an indeterminism linked to the non-existence of a specific function $\tau(t)$ and no longer to a conceptual dualism, as in the standard interpretation. In this model we still have a corpuscular and a wave point of view, the latter analogous to the standard one. However, there is no incompatibility between them, but the wave point of view is necessary due to the described constitutional insufficiency of the corpuscular description.

{In this section we have outlined the two-time model, highlighting its basic properties. In the next one we will study the free particle}.

\section{\emph{Zitterbewegung }in the new model}
Remarkably, a \emph{Zitterbewegung}
motion in the $\tau$ variable emerges after three integrations by studying the free particle in a fixed direction.

Here, basically, we will follow the deduction made in \cite{40}; but, thanks to the more general form of eq. (24), obtained using eq. (22), we will get a simpler and more rigorous derivation.

The classical motion equation (20) just informs us that $\vec{p}_{c}$ is constant. All equations (7) through (14) found in section II are valid.
Again, eq. (21) can be rewritten:
\begin{equation}
\frac{\partial F}{\partial t}+\vec{v_c}\cdot\vec{\nabla}F=\frac{{\rm{d}}F}{{\rm{d}}t}=0,
\end{equation}
being $F$ any function of any-order derivative of $R$ over $t$ or space. In particular it holds for $V_Q$ and $H$: they are therefore fields that travel in space, although  with zero total derivative with respect to $t$-time.

We will begin by studying only the classical rectilinear trajectory, called $\vec{x}=x\hat{v_c}$. Here, eq. (25) becomes the equation of a progressive wave:
\begin{equation}
\frac{\partial F}{\partial t}+{v_c}\dot{F}=0,
\end{equation}
with a dot denoting the derivative with respect to \emph{x}. Sure, deriving again, we obtain the d'Alembert equation:
\begin{equation}
\frac{\partial^2 F}{\partial t^2}-{v_c^2}\ddot{F}=0.
\end{equation}
So, $V_{Q}$ and \emph{H} are progressive waves, i.e. they follow the particle in its $t$-motion. 

Using the above equations for \emph{F=R}, eqs. (11) and (14) can be rewritten as:
\begin{equation}
H=\pm\sqrt{m^{2}\gamma_{c}^{2}c^{4}-\hbar{}^{2}\frac{c^{2}}{\gamma_{c}^{2}}\frac{\ddot{R}}{R}},
\end{equation}
\begin{equation}
v_{c}\,\frac{\dot{R}}{R}\,(m\,\gamma_{c}\,c^{2}-H)+\frac{1}{2}\frac{\partial H}{\partial t}=0.
\end{equation}

In the following we assume \emph{$m\gamma_{c}\neq0$}, \emph{ $v_{c}\neq0$}, $\dot{R}\neq0$ and limit to consider only non-negative
values for the Hamiltonian.
Substituting eq. (28) into eq. (29), we get:
\begin{equation}
v_{c}\,\frac{\dot{R}}{R}\,(\sqrt{\beta}-\beta)=-\frac{1}{4}\frac{\partial\beta}{\partial t},
\end{equation}
where $\beta\equiv\frac{H^{2}}{m^{2}\gamma_{c}^{2}c^{4}}=1-\lambda^{2}\,\frac{\ddot{R}}{R}$, with  $\lambda^{2}\equiv\frac{\hbar{}^{2}}{m^{2}c^{2}\gamma_{c}^{4}}$. Using eq. (26) for $F=\beta$, one finds:
\begin{equation}
\frac{\dot{R\,}}{R}\,=\frac{1}{4}\,\frac{\dot{\beta}}{\sqrt{\beta}-\beta}.
\end{equation}
Multiplying by \emph{dx} and integrating member by member we can obtain:
\begin{equation}
1+\frac{c_{1}}{R^{2}}=\sqrt{\beta},
\end{equation}
with $c_{1}$ arbitrary constant. So:
\begin{equation}
H=m\gamma_{c}c^{2}\,(1+\frac{c_{1}}{R^{2}}),
\end{equation}
and the Quantum Potential is $V_{Q}=m\gamma_{c}c^{2}\,\frac{c_{1}}{R^{2}}$. Compatibility with the stationary non-relativistic case requires
$c_{1}>0$. By squaring eq. (32) one finds:
\begin{equation}
(1+\frac{c_{1}}{R^{2}})^{2}=1-\lambda^{2}\,\frac{\ddot{R}}{R}.
\end{equation}
Recalling that \emph{$R>0$}, a first integration provides:
\begin{equation}
\dot{R}=\pm\frac{1}{\lambda}\,\sqrt{\frac{c_{1}^{2}}{R^{2}}-4c_{1}{\rm{ln}}\,R+c_{2}},
\end{equation}
with $c_{2}$ arbitrary constant. We look for a solution where \emph{R}
has a maximum, called $R_{M}$, at some point \emph{$x_M$}, having
in eq. (35) sign - for \emph{$x>x_M$} and + for \emph{$x<x_M$}. From $\dot{R}(x_{M})=0$ we determine $c_{2}$, getting:
\begin{equation}
\dot{R}=\pm\frac{\sqrt{c_{1}}}{\lambda}\,\sqrt{\frac{c_{1}}{R^{2}}-\frac{c_{1}}{R_{M}^{2}}+4\,{\rm{ln}}\,\frac{R_{M}}{R}}.
\end{equation}
Placing $f\equiv\frac{c_{1}}{R_{M}^{2}}>0$, the Quantum Potential is rewritten:
\begin{equation}
V_{Q}=m\gamma_{c}c^{2}f\,\frac{R_{M}^{2}}{R^{2}},
\end{equation}
and its minimum value is in $x_{M}$: $V_{Qm}=m\gamma_{c}c^{2}f$. \emph{R} is determined by integrating eq. (36):
\begin{equation}
\int\frac{{\rm{d}}R}{\sqrt{f(\frac{R_{M}^{2}}{R^{2}}-1)+4\,{\rm{ln}}\,\frac{R_{M}}{R}}}=-\frac{R_{M}\sqrt{f}}{\lambda}\mid x-x_{M}\mid+c_{3},
\end{equation}
which cannot be simplified by means of standard functions.

In a small neighborhood of the maximum, i.e. for $R\rightarrow R_{M}$, eq. (38) gives for \emph{R} a quadratic dependence on $x-x_{M}$.
Indeed, here the rooting at first member is approximated by $2(f+2)(1-\frac{R}{R_{M}})$, so by integrating we have:
\begin{equation}
\sqrt{1-\frac{R}{R_{M}}}\simeq\frac{\sqrt{f(f+2)}}{\sqrt{2}\lambda}\,\mid x-x_{M}\mid+c_{4},
\end{equation}
where $c_{4}=0$ by imposing $R(x_{M})=R_{M}$. Hence:
\begin{equation}
R\simeq R_{M}(1-\frac{f(f+2)}{2\lambda^{2}}(x-x_{M})^{2}),
\end{equation}
\begin{eqnarray}
\frac{1}{R^{2}}\simeq\frac{1}{R_{M}^{2}}\frac{1}{1-\frac{f(f+2)}{\lambda^{2}}(x-x_{M})^{2}}\simeq\nonumber\\
\simeq\frac{1}{R_{M}^{2}}(1+\frac{f(f+2)}{\lambda^{2}}(x-x_{M})^{2}).
\end{eqnarray}
Then, from eq. (37) the Quantum Potential around $x_{M}$ can
be approximated by the potential of a harmonic oscillator:
\begin{equation}
V_{Q}\simeq m\gamma_{c}c^{2}f\,(1+\frac{f(f+2)}{\lambda^{2}}(x-x_{M})^{2}).
\end{equation}
By replacing it into the $\tau$-motion equation (24) and imposing $x_M=X(t,0)=X_{t_0}+X_{\tau_0}+v_c\,t$, we ultimately get the following solutions:
\begin{equation}
X(t,\tau)\simeq X_{t_0}+X_{\tau_0}+v_c\,t+\Delta X\,{\rm{sin}}(\frac{c}{\lambda}\,f\sqrt{2(f+2)}\tau),
\end{equation}
\begin{equation}
v_{i_x}(\tau)\simeq \Delta X\,\frac{c}{\lambda}\,f\sqrt{2(f+2)}\,{\rm{cos}}(\frac{c}{\lambda}\,f\sqrt{2(f+2)}\tau),
\end{equation}
where, according to point \emph{1} of the theory,  the amplitude $\Delta X$ represents the positional indeterminacy in $x$ of the particle. To get amplitude and angular frequency of the standard \emph{Zitterbewegung}, i.e. $\frac{\hbar}{2m\gamma_{c}c}$ and $\frac{2m\gamma_{c}c{}^{2}}{\hbar}$, it is enough to set $\gamma_{c}\,f\sqrt{2(f+2)}=2$, and $v_{i_{x_{MAX}}}=c$. However, by imposing only $v_{i_{x_{MAX}}}=c$, we obtain $\Delta X=\frac{\lambda}{f\sqrt{2(f+2)}}$; as for \emph{f}, we will simply assume that it is of the order of unity. Then, in the model under consideration, the uncertainty principle is: 
\begin{equation}
\frac{\lambda}{f\sqrt{2(f+2)}}\times m\gamma_{c}c=\frac{\hbar}{f\sqrt{2(f+2)}\gamma_{c}}\sim\frac{\hbar}{\gamma_{c}},
\end{equation}
with an unexpected Lorentz factor to divide. Looking at eq. (28) this seems at all correct, because there is not only a $\gamma_{c}$ which increases
the relativistic mass but also \emph{another} $\gamma_{c}$ which decreases
the influence of $\hbar$. At ultrarelativistic speeds, the quantum positional uncertainty in direction of motion should be negligible.

{Before continuing the analysis it is opportune to comment on what we would have found if we had assumed the Special Relativity be valid for $\tau$-motion. The study of a relativistic oscillator \cite{66}, provides the following result in our case: 
\begin{equation}
    \Delta X=\frac{c\sqrt{2}\sqrt{1-\sqrt{1-\beta^2}}}{\omega_0(1-\beta^2)^\frac{1}{4}},
\end{equation}
where $\omega_0$ is the non-relativistic pulsation and $\beta=\frac{v_{i_{x_{MAX}}}}{c}$. But this is incompatible with our model: by imposing $v_{i_{x_{MAX}}}=c$, we obtain $\Delta X\rightarrow \infty$, therefore not being able to represent the minimum positional uncertainty. }

The amplitude of the $\tau$-oscillation, i.e. the quantum positional uncertainty, is inversely proportional to the rest mass. Then, for a particle of one picogram we find it about $15$ orders of magnitude smaller than for an electron. However, this is not the only reason that explains the evanescence of the $\tau$-motions for macroscopic bodies, as we will show in the next section. 

The considered $\Delta X$ represents the minimum positional uncertainty of the particle; only by means of a wave packet description is it possible to take into account larger positional uncertainties (and smaller momentum uncertainties).

For $x=X$ in eq. (42), we get the dependence of $V_{Q}$ on $\tau$: it oscillates, and so does $H$, with an amplitude $\frac{1}{2}m\gamma_{c}c^{2}$; this value represents so
the quantum uncertainty in a measure of energy, for a free particle
with minimal positional uncertainty. Of course, multiplying by
half-period of oscillation - which represents the quantum uncertainty in a measure of $t$-time - we find again the uncertainty principle:
\begin{equation}
\frac{1}{2}m\gamma_{c}c^{2}\times\frac{\pi\lambda}{f\sqrt{2(f+2)\,}c}=\frac{\pi\hbar}{2\,f\sqrt{2(f+2)}\gamma_{c}}\sim\frac{\hbar}{\gamma_{c}}.
\end{equation}

The purely quantum total energy $E_{q_x}\equiv V_{Q}+\frac{1}{2}m\gamma_{c}v_{i_x}^{2}$
is independent of $x$: 
\begin{equation}
E_{q}=V_{Qm}+\frac{1}{2}m\gamma_{c}c{}^{2}=m\gamma_{c}c^{2}(f+\frac{1}{2}),
\end{equation}
and it also is  the maximum value of $V_{Q}$, corresponding to the minimum value of \emph{R}: $R_{m}=\frac{R_{M}}{\sqrt{1+\frac{1}{2f}}}$. Certainly, a normalization of $R{}^{2}$ (not exact, for the non-integrability
of eq. (38)) could determine $R_{M}$.

As an important result it was therefore found not simply that the Quantum Potential is never negligible, but that its maximum value, which represents the total purely quantum energy, is at least of the order of $m\gamma_{c}c^{2}$, regardless of the value of $f$. Nonetheless, the $\tau$-motions can be negligible in amplitude, as happens for a macroscopic body. It is also surprising that, although the existence of $V_Q$ is due to $\hbar$, its maximum and minimum values do not depend on it: only the frequency of its oscillation does.

It is noteworthy that \emph{Zitterbewegung} in $\tau$-time is able to justify the physical origin of the phase to be associated with each path in Feynman's formulation of Quantum Mechanics. The approach has been specified in \cite{67}, through a particle's oscillation forward and backward in \emph{t}-time. In this model, that type of motion arises spontaneously: it is the found \emph{Zitterbewegung} in $\tau$, represented by the wavefunction  extra phase factor ${\rm{e}}^{\frac{{\rm{i}}}{\hbar}S_\tau}$.

{We have so found the \emph{Zitterbewegung} in direction of motion; in the following three subsections we generalize and complete its study.
}

\subsection{\emph{Zitterbewegung} in arbitrary direction}
{With respect to an arbitrary spatial direction $\hat{s}\neq \hat v_c$, let's consider the polar coordinate system ($s$, $\theta$, $\phi$), where $\theta \neq 0$ is the angle between $\hat{s}$ and $\hat v_c$ and $\phi$ the azimuth (fig. 1). \begin{figure} [h!] \includegraphics[width=5.3cm, height=5cm]{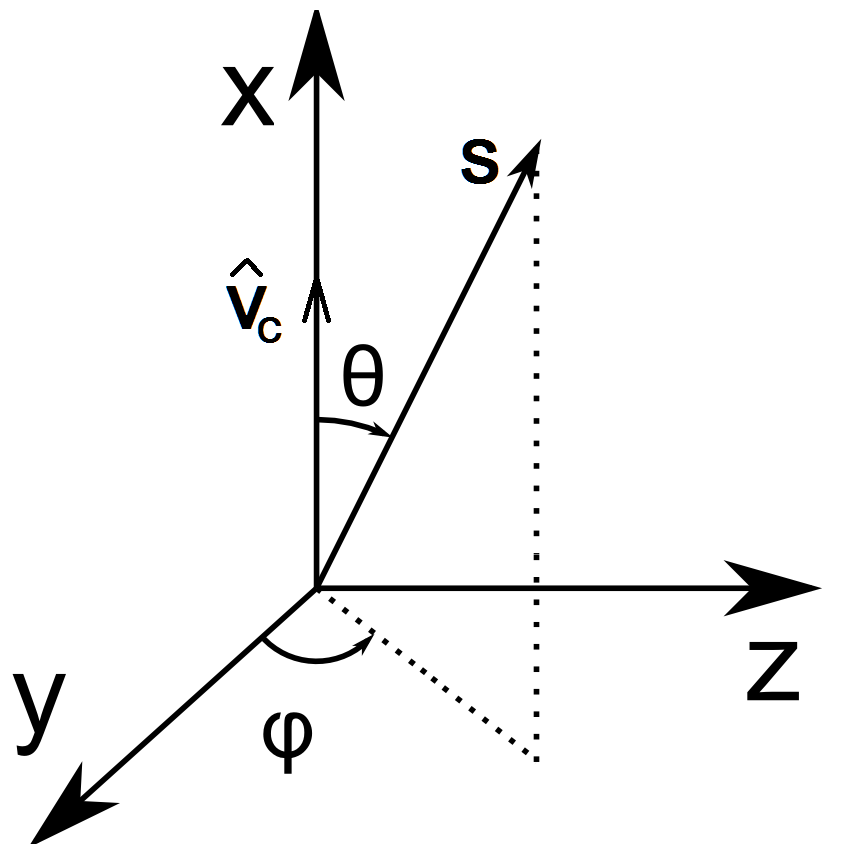}
\caption{Polar coordinate system for studying the \emph{Zitterbewegung} in an arbitrary direction $s$.}
\end{figure}
Having as our goal to find the $\tau$-motion along $s$ by eq. (24), $\theta$ and $\phi$ must be considered constant, so the polar expressions of $\vec{\nabla} R$ and $\nabla^2 R$ reduce to:
\begin{equation}
\vec{\nabla} R=\frac{{\rm{d}}R}{{\rm{d}}s}\hat{s},\quad\quad \nabla^2 R=\frac{2}{s}\frac{{\rm{d}}R}{{\rm{d}}s}+\frac{{\rm{d}}^2R}{{\rm{d}}s^2}.
\end{equation}
Substituting the first one into eq. (25) and differentiating on $t$ gives:
\begin{equation}
\frac{\partial^2 R}{\partial t^2}-{v_{c_s}^2}R''=0,
\end{equation}
where $v_{c_s}=v_c\ {\rm{cos}}\ \theta$ and the apostrophe indicates derivative over $s$.
Substituting eq. (50) and the second of eqs. (49) into eq. (11) and rooting is found:}

{\begin{equation}
H=\sqrt{m^{2}\gamma_{c}^{2}c^{4}-\hbar^{2}\frac{c^{2}}{\gamma_{c_s}^{2}}\frac{R''}{R}-\frac{2\hbar^2c^2}{sR}R'}\ ,\ \gamma_{c_s}\neq \gamma_c,
\end{equation}
while eq. (14) is generalized by:
\begin{equation}
v_{c_s}\,\frac{R'}{R}\,(m\,\gamma_{c}\,c^{2}-H)+\frac{1}{2}\frac{\partial H}{\partial t}=0.
\end{equation} Now, remembering that, in order to integrate eq. (38), we had to limit ourselves to small neighborhoods of the maximum $R_M$, we can neglect the last addend of eq. (51) and thus lead back to the already studied eqs. (28) and (29).}
Placing: $\lambda^{2}\equiv\frac{\hbar{}^{2}}{m^{2}c^{2}\gamma_{c}^{2}\gamma_{c_s}^{2}}$, we so get back the same equations found before. Expressions of the Hamiltonian
and Quantum Potential, given by (33) and (37), do not change,
not even the amplitude of their oscillations. By imposing the $\tau$-motion equation,
we obtain an intrinsic harmonic motion described again by (43) and (44), but with the new value of $\lambda$.

{The different value of $\lambda$ therefore causes a dependence of the oscillation amplitude and frequency on the spatial direction: this is precisely the relativistic anisotropy $v_{i_s}\neq v_{i_u}$ that we anticipated in section III.} Remarkably, this is reflected in a \emph{spatial anisotropy of the uncertainty principle}, where we have the presence of $\gamma_{c_s}$
instead of $\gamma_{c}$ to divide:
\begin{equation}
\Delta s\cdot\Delta p_{s}\sim\frac{\hbar}{\gamma_{c_s}}.
\end{equation}
In particular, for \emph{$\hat{s}$} tending to be perpendicular
to\emph{ $\hat{v}_{c}$}, $\gamma_{c_s}\rightarrow1$ and the standard uncertainty
principle, independent of speed, is re-established. However, the case
of \emph{exact} perpendicularity is singular: for $v_{c_s}=0$ eq.
(52) implies\emph{ $H=const\Rightarrow V_{Q}= const\Rightarrow{v}_{i_s}= const$}, instead  of being oscillatory.
\subsection{Non-relativistic limit. Singularity in $v_{c}=0$}
For the non-relativistic ($v_{c}\ll c$) and $v_{c}\rightarrow0$
cases, it is sufficient to approximate $\gamma_{c}$ in the previous
equations (respectively, with $1+\frac{v_{c}^{2}}{2c^{2}}$ and 1),
\emph{obtaining still high frequency oscillatory motions in} $\tau$. \emph{These results cannot be obtained starting from the standard Schr{\"o}dinger
equation} because from it the
non-relativistic approximation of eq. (52) cannot be deduced. This happens since the explicit dependence of $R$ on $t$, causes the Klein-Gordon equation to no longer tend to the Schr{\"o}dinger equation.

In $v_{c}=0$ there is another singularity: as before, we
get by eq. (52) ${v}_{i_s}=const$ for any $\hat s$, contrary to the case $v_{c}\rightarrow0$. 
\subsection{Standard \emph{Zitterbewegung}}
Interpretating  the generic oscillatory motion in $\tau$ as occurring in \emph{t}, the standard \emph{Zitterbewegung} is re-obtained.
 In the standard view, $R$ is considered time independent, so
the second side of eq. (10) is just $-\frac{H^{2}}{\hbar{}^{2}}$, with $H=const$. In the new theory this means demanding: $\frac{1}{R}\frac{\partial^{2}R}{\partial t^{2}}-2\,\frac{\rm{i}}{\hbar}\,\frac{H}{R}\frac{\partial R}{\partial t}=0$, that provides $R\propto\,{\rm{e}}^{\frac{2{\rm{i}}}{\hbar}H\,t}$ and hence $V_{Q}\propto {\rm{e}}^{-\frac{4{\rm{i}}}{\hbar}H\,t}$. Then, using eqs. (22) for $V_{Q}$ and (24) (both as for \emph{t}-time), after having conveniently determined the arbitrary constants, one can get: $v_{i_s}\propto c\,{\rm{e}}^{-\frac{2{\rm{i}}}{\hbar}H\,t}$, as found in the Heisenberg \emph{picture}.
\section{Interpretation of the double-slit experiment}
The described model, without questioning the traditional conservation laws, also adopts the standard evolution of the particle wavefunction through the Klein-Gordon equation. The introduction of the independent time parameter, which is hidden, basically provides a deterministic explanation of the uncertainty principle. We have seen that the latter changes in the direction of motion only at relativistic speeds. Excluding for now this case, which will be discussed in the next section, it can be stated that the empirical predictions of standard Quantum Mechanics are fully respected. For example, no difference is foreseen for the kinetic theory of gases, whose standard quantum model is obtained by discretizing the phase space into volumes $h^\emph{3}$. However, the model is able to resolve the paradoxical dualism that afflicts the standard interpretation, by explaining deterministically the collapse of the wave packet after a measurement. We will now detail how the theory works in the case of the most famous quantum experience: the double-slit experiment.

The standard wave packet description, compliant the de Broglie relation, is still empirically justified on the basis of the physical indeterminations of the sent particle. At the initial instant, we assume for simplicity that the particle has the minimum positional uncertainty previously found, which we call $\Delta s_{0}$. During motion, as long as the energy of the particle does not change, the growth of the positional uncertainty, obtainable by the Liouville's theorem, continues to be the standard one:  $\Delta s=\sqrt{\Delta s_{0}^{2}+\frac{\Delta p_{s_0}^{2}}{m^{2}}\,t^{2}}$.
At a fixed instant $\bar t$, $\Delta s$ represents the radius of a sphere-like figure within which the particle, due to the omnidirectional $\tau$-motions, is spread; in other words, at instant $\bar t$ it is \emph{simultaneously present in every point of this volume}. We highlight that:
\begin{enumerate}
    \item This simultaneity is not in conflict with Special Relativity, because the generic $\tau$-motion itself, as long as the energy of the particle does not change, does not constitute a $t$-\emph{event}: $\bar t$ is {\emph{frozen}} in any reference system.
  \item Nothing prevents these infinite \emph{self-particles}  from $\tau$-interfering with each other in all possible physical ways: each kind of interacting particles (photons, vector bosons, etc.) oscillates itself in $\tau$-time according to the same laws found. Mutual exchanges of particles correspond to what in the standard point of view is described as emission and reabsorption of virtual particles, obtaining excellent experimental verifications.
\item In addition to $\tau$-interacting like particles, self-particles $\tau$-interfere like waves, exactly as predicted by the standard interpretation, according to the Feynman description \cite{68}. The model simply specifies that the \emph{different histories} are \emph{actually  realized} in time $\tau$. This description is in conflict neither with the interactions just discussed nor with the corpuscular description given by equations (20) and (24). Due to the constitutional limitations of the latter, the effective trajectory followed by the particle is everywhere unknowable and the wave description is then  necessary. When the particle reaches the wall, diffraction is observed if the sphere-like volume includes the two slits (fig. 2).
    
\end{enumerate}

\begin{figure} [h]

\includegraphics[width=8.6cm, height=6.7
cm]{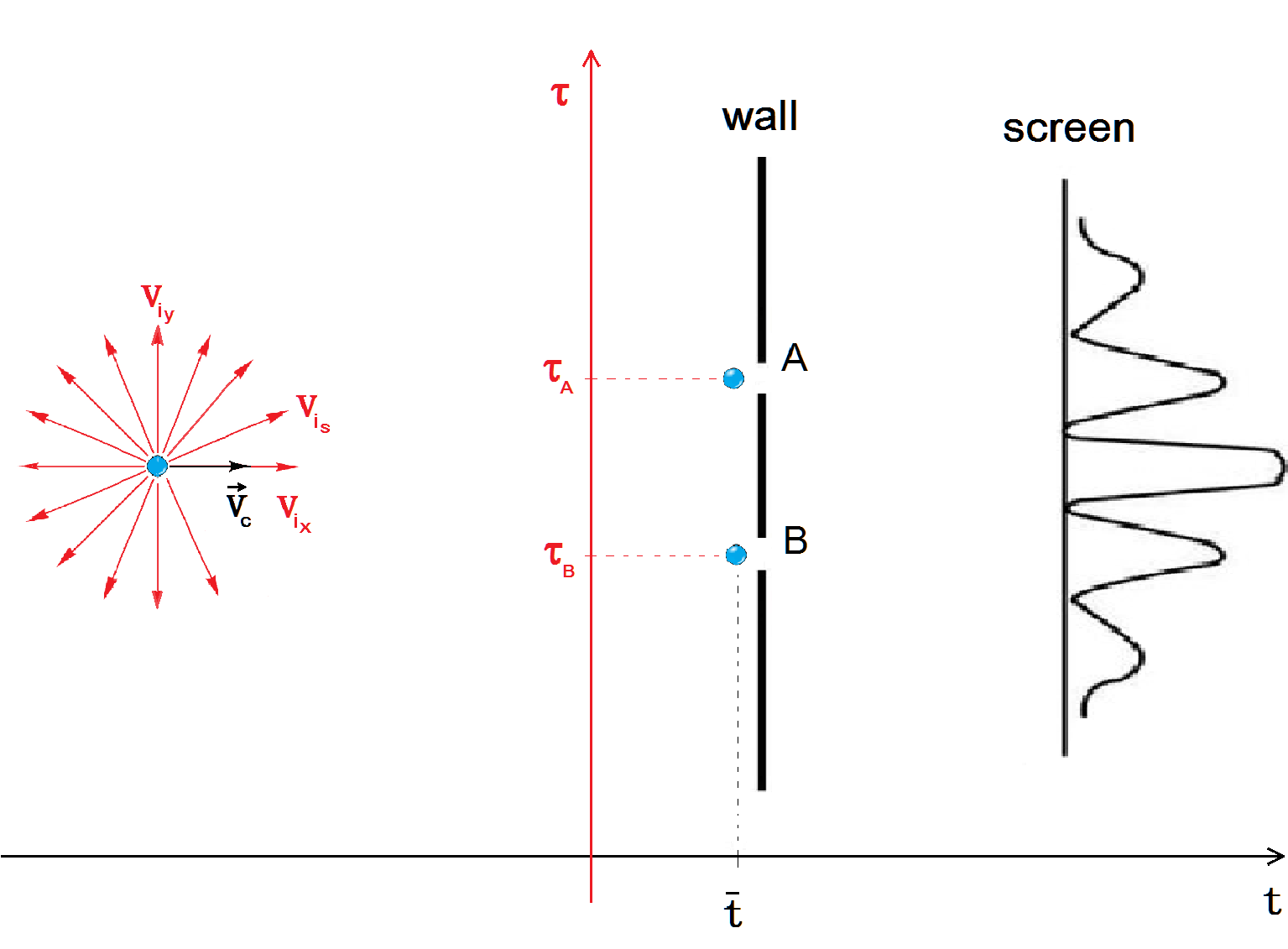}
\caption{Interpretation of the two slits experience according to the Two-Time Model: with respect to the hidden time $\tau$, independent of the normal time $t$, the particle reaches the slits at two different instants; but with respect to $t$ its presence in them is simultaneous.}
\end{figure}

Note that zig-zag trajectories which violate the conservation of $t$-momentum and yet admitted as legitimate Feynman paths, can now be \emph{properly physical}, being able to be obtained by composition of $\vec p_c$ with one or more $\vec p_{i_s}$.

With respect to time $t$ (that is, relative to an observer), the particle travels \emph{at once} all the possible paths between the initial point and the slits, provided that they are \emph{included in the volume swept by the sphere-like figure}. In this, therefore, the model differs from the standard \emph{sum of histories }interpretation, where there is no limitation for the paths.

\subsection{Collapses}

Areas where the self-interferences are
destructive constitute "shadow zones" of the sphere-like volume, where from a corpuscular point of view, the Quantum Potential pushes the self-particles elsewhere. Dually, there will also be areas where the interference is constructive.

The apparent observational "collapse" is an extreme case of destructive interference that occurs after a energy variation of the particle for collision (more generally, interaction) with an external particle or field: what happens is that the packet wavefunctions located in the impact area abruptly change their phase proportional to $S_t$ and then interfere destructively (statistically) with those located elsewhere. Hence the packet reconfigures itself - instantaneously, with respect to $t$-time - by centering itself in the impact area and restricting its dispersion. From a corpuscular point of view, the Quantum Potential no
longer allows the particle to $\tau$-move away from the
proximity of the impact.

So, a non-zero energetic interaction with an external particle or field constitutes an \emph{observable} (but not necessarily observed) $t$-\emph{event}, capable of dramatically restricting $\Delta s$, its minimum value being of the Compton length order. From a macroscopic
view, the intrinsic movements stop. Observation itself
has no special role: this happens regardless of whether
the external particle is from a detector or residual air.

Consistent with this description, after a perfectly elastic collision with a massive wall (it is also the case of a photon hitting a perfectly reflective mirror) the particle does not reduce its delocalization: no energy is transferred to the wall and the $\tau$-motions amplitude does not collapse.

This picture also indicates another fundamental reason that generally prevents a macroscopic body from spread out in space through the intrinsic motions: compared to a particle, it is vastly more likely that it interacts energetically with other matter or fields. Indeed, even a small  wave phase variation  is capable of making it collapse in the way described, canceling out significant quantum uncertainties.

For what has been said, the paradox that the particle, being $t$-simultaneously in every point of a certain macroscopic volume, violates the conservation of mass and of number of particles is only \emph{apparent}: a particle search experiment will always detect only \emph{one} particle. The total purely quantum energy distributed in the sphere-like volume at any $t$-instant is indeed infinite, but it is beyond any kind of direct experimentation. Actually it is \emph{virtual}, in the same sense used in standard view. Indeed, the existence of the hidden time $\tau$ is capable of explaining why the increasingly larger energies predicted by entering the infinitesimal are real (as experiments excellently confirm) but never directly observed.
\section{Discussion}
In Feynman's (however \emph{standard}) formulation of Quantum Mechanics, when a particle moves from one point to another, it is \emph{as if} it followed \emph{at once} all possible paths between the two points. In this article we have begun to explore a model in which, by abandoning the \emph{"as if"}, the particle \emph{really} does so. For this to be possible without contradiction, it is necessary to add a new time parameter, totally independent of the commonly experienced one. Called it $\tau$, the generic wavefunction of the particle is represented by eq. (18). The equations that are obtained by differentiating the latter with respect to time $t$ consist just of a relativistic generalization of the classical Bohmian ones. Conversely, by assuming for the generic  $\tau$-motion an appropriate wave  equation, new laws of motion, which must replace that one of Bohm's original theory, derive. The $\tau$-motions are governed by the Quantum Potential and spread
out the particle in a region of space that depends on the conditions of the physical
surrounding. For a free particle, a more general \emph{Zitterbewegung}
motion has been found, having minimum amplitude of the order of the Compton length. In presence of a $V\neq0$, the Quantum Potential is modified through eqs. (11) and (12) and the spreading through eq. (24); in the case of an atomic electron, for example, the $\tau$-motions justify the instantaneous dispersion of the electron in the orbit (we intend to show this in a subsequent publication).

In the model, the quantum uncertainties and the  uncertainty principle itself have a deterministic explanation: they are due to the $\tau$-movements. For the latter, the Special Relativity cannot hold, but since they are instantaneous
with respect to \emph{t}-time in any reference system, it is impossible to
detect violations of the Lorentz covariance.

We have to emphasize that, since \emph{$m\gamma_{c}\neq0$} is the only hypothesis made about the particle, all the conclusions obtained - including the characteristics of the oscillatory motions in $\tau$ - continue to be valid for photons (for which, on the other hand, quantum diffraction figures analogous to those of any other particle are found).

A question concerns the determination of the parameter \emph{f}. For example, by assuming \emph{f = 1/2}, one gets, in any direction,  a total purely quantum energy $m\gamma_{c}c^{2}$, equal to that of the particle. However we find no arguments to suggest whether this is \emph{the} (or \emph{an})  appropriate value. It's clear that the determination of \emph{f} should derive by rigorous considerations and it is not excluded that it may depend on the type of particle and/or its properties here not considered.

The evanescence of the uncertainty principle in direction of motion for   $v_{c}\rightarrow c$, is recognizable just by observing
the general eq. (51): there, a Lorentz factor - equal to 1 only in direction tending to be perpendicular to the
rectilinear classic motion - directly reduces the influence of $\hbar$.
Then, the spreading region of the particle should be contracted in
direction of motion.
Believing that the high energy available in today's accelerators should
have already detected deviations based on eq. (53), would be quite
hasty. It is necessary to consider not only that high speeds make
localization measurements hard but above all that simultaneous measurements
of complementary observables are prohibited by the same principle
of causality in  Special Relativity. These and several other problems
make it very difficult to experimentally verify the uncertainty principle \cite{69,70}.

Compared to spacetime structures planned by the current most accredited theories, introduction of a second time variable is a more radical but also mathematically simpler criterion which, neither logically nor physically, differs in essence from the introduction of an extra spatial dimension. In addition to the explanation of the \emph{Zitterbewegung}, it predicts new physical effects, permitting, inside the Feynman's formulation, to clarify the phase of the paths (redefinable as \emph{real physical} trajectories, although not always observable) and hopefully to represent the sum over histories as a properly mathematical integral.

In the explored theory, the non-locality and the apparently paradoxical aspects of Quantum Mechanics are just the effect of the existence of an extra unobservable time dimension. By traveling in a hidden time, you can deterministically explain why a particle or photon is able to
 interact instantly with more slits in a diffraction, to run across multiple paths between two points, to admit
entanglements with other arbitrarily distant particles, to emit and reabsorb many species of virtual particles... All established facts.
\section*{Appendix}

{In this section we provide a justification of the peculiar wave equation (22), necessary to deduce the $\tau$-motion law. First of all, taking advantage of the reported generality, it is convenient to refer it to $R^2$:
\begin{equation}
\frac{\partial R^2}{\partial \tau}-v_{i_s}{\frac{\partial R^2}{\partial s}}=0.
\end{equation}
 When one refers this motion to an oscillation, one realizes that $R^2$ is in \emph{antiphase} - i.e. shifted by half a period - with respect to the particle moving with speed $v_{i_s}$. This "pursuing" reminds the way the \emph{Zitterbewegung} is interpreted in the standard interpretation: the particle "chases" its antiparticle (and vice versa) which, according to the uncertainty principle, is created for lengths $\sim\lambda_c$ (see e.g. \cite{71}). However, our model makes no reference to antimatter, admitting only positive values for energy. In fact, we  are able to justify this equation by referring exclusively to the meaning of the physical quantities involved. According to our model, the operation of finding the particle in a unit volume (whose probability is $R^2$) must specify a $\bar t$-event. Then, for an amplitude $\sim\lambda_c$, $\bar t$ must have an uncertainty of about half a period of the intrinsic oscillation of the particle, as we have shown. And this \emph{effectively means} that the particle must perform half an oscillation - the cause of uncertainty - before the measurement is achievable, ending up in antiphase with $R^2$.}
 \section{Acknowledgments}

 I am indebted to H. Nikolić and A. J. Silenko for their support. I also especially thank the reviewer of the manuscript for encouraging me to improve it significantly. Besides, the promoters and developers  of \emph{WolframAlpha}
to make free available on the Web their computational engine.

\providecommand{\noopsort}[1]{}\providecommand{\singleletter}[1]{#1}%


\end{document}